\pdfoutput=1
\documentclass[useAMS,usenatbib]{mn2e}
\usepackage{graphicx}
\usepackage{amssymb}
\usepackage{lscape}
\usepackage{rotating}
\usepackage{subfigure}
\usepackage{amsmath}
\usepackage{textcmds}

\def\deg{\ifmmode^\circ\else$^\circ$\fi}

\def\Q{\ifmmode\mathcal{Q}\else$\mathcal{Q}$\fi}
\def\Mach{\ifmmode\mathcal{M}\else$\mathcal{M}$\fi}
\title[Possible twisted filaments in Lynds Bright Nebulae]
{Lynds Bright Nebulae: Sites of possible twisted filaments and ongoing star formation}
\author[L.~K. Dewangan et al.]
{L.~K. Dewangan$^{1}$\thanks{lokeshd@prl.res.in}, J.~S. Dhanya$^{2}$, N.~K. Bhadari$^{1,3}$, D.~K. Ojha$^{4}$, and T. Baug$^{5}$\\
$^{1}$Physical Research Laboratory, Navrangpura, Ahmedabad - 380 009, India.\\
$^2$Malaviya National Institute of Technology (MNIT), Jaipur 302017, Rajasthan, India.\\
$^3$Indian Institute of Technology Gandhinagar Palaj, Gandhinagar, 382355, India.\\
$^4$Department of Astronomy and Astrophysics, Tata Institute of Fundamental Research, Homi Bhabha Road, 
Mumbai 400 005, India.\\
$^5$Satyendra Nath Bose National Centre for Basic Sciences, Block-JD, Sector-III, Salt Lake, Kolkata-700 106, India.\\
}
\begin{document}

\date{ }

\pagerange{\pageref{firstpage}--\pageref{lastpage}} \pubyear{2020}

\maketitle

\label{firstpage}

\begin{abstract}
The paper presents an analysis of multi-wavelength data of two Lynds Bright Nebulae (LBN), LBN 140.07+01.64 and LBN 140.77$-$1.42. 
The 1420 MHz continuum map reveals an extended Y-shaped feature (linear extent $\sim$3$\degr$.7), which consists of a linear part and a V-like structure. The sites LBN 140.07+01.64 and AFGL 437 are located toward the opposite sides of the V-like structure, and LBN 140.77$-$1.42 is spatially seen toward the linear part. Infrared-excess sources are traced toward the entire Y-feature, suggesting star formation activities. Infrared and sub-millimeter images show the presence of at least two large-scale dust filaments extended toward the LBN sources. 
The {\it Herschel} maps, which are available only toward the northern and central parts of the Y-feature, display the presence of higher column density ($\ge$ 2.4 $\times$10$^{21}$~cm$^{-2}$) of materials toward the filaments. 
Using the $^{12}$CO(1--0) line data, the distribution of molecular gas at [$-$42.7, $-$34.4] km s$^{-1}$ traces the cloud associated with the Y-feature, and confirms the existence of filaments. The large-scale filaments appear to be possibly spatially twisted. There is a hint of an oscillatory-like velocity pattern along both the filaments, favouring their proposed twisted nature. It is the first study showing the possible twisting of filaments, which is more prominent in the northern and central parts of the Y-feature. This possible twisting/coupling of the large-scale filaments appears to be responsible for the observed star formation (including known OB-stars). 
The proposed physical process and the energetics of OB-stars together seem to explain the origin of the ionized Y-feature.
\end{abstract}
\section{Introduction}
\label{sec:intro}
In the last decade, the study of infrared (IR) and sub-millimeter (sub-mm) observations revealed that filaments are common structures in low-mass and high-mass star-forming regions and their role in the star formation processes has been evident \citep[e.g.,][]{myers09,andre10,dewangan15,dewangan17b,dewangan18,dewangan19,dewangan20x,dewangan21,Tige+2017,motte18,morales19,kumar20}. In this context, one can study the signatures of the convergence of filaments toward the compact and dense hub, the intersection/merging of filaments, and the collisions of filaments, which can explain the ongoing physical processes in star-forming regions. 
In such events/scenarios, the highest column density can be observed in the interaction zones of filaments. Additionally, a filament braid, where continuous multiple filaments are twisted, could also be considered as one of the interesting star-forming sites. In such configuration, multiple common zones with high column densities are expected. 
However, in the literature, we do not find any promising star-forming sites where the twisting of filaments is investigated. 
In this relation, the target of our present paper is two Lynds Bright Nebulae (LBN), LBN 140.07+01.64 and LBN 140.77$-$1.42 \citep{Lynds1965,Green1989b,Karr2003}.

LBN 140.77$-$1.42 has been investigated as a linear radio ridge \citep[extent $\sim$2$\degr$;][]{Green1989b}, while 
LBN 140.07+1.64 has a more diffuse appearance in the radio 1420 MHz continuum map \citep{Karr2003}.
The radio emission detected toward both the LBN sources was found to be thermal in nature \citep{Green1989b,Karr2003}. 
The molecular gas at V$_{\rm lsr}$ = [$-$45.2, $-$32.8] km s$^{-1}$ and the dust emission were traced toward the LBN sources \citep{heyer01,Karr2003}. 
Using the molecular line data (i.e., $^{12}$CO(1--0), $^{13}$CO(1--0), and C$^{18}$O(1--0)) from the Purple Mountain Observatory Delingha (PMODLH) 
13.7 m telescope, \citet{Du2017} also studied the molecular emission toward an area containing both the LBN sources, which was designated as ``Grand Canal" (see Figure~10 in their paper). The Grand Canal was reported as a giant filamentary molecular cloud (GFMC) in the
Perseus Arm \citep{Du2017}, and was divided into three parts \citep[see also IDs 50--52 in][]{digel96}. 
\citet{Karr2003} identified young stellar objects (YSOs) based on the IRAS color conditions, which were found toward the Grand Canal hosting the LBN sources (see Figure~3 in their paper). 

Figure~\ref{fig1}a shows the radio 1420 MHz continuum map of an area containing LBN 140.07+01.64 and LBN 140.77$-$1.42.
The positions of a few known star-forming sites/H\,{\sc ii} regions (i.e., IRAS 03035+5819/AFGL~437, AFGL~5090, IRAS 03101+5821, 
and IRAS 03063+5735) 
are also indicated in Figure~\ref{fig1}a. The positions of molecular clumps \citep[from][]{heyer01} are also overlaid on the 1420 MHz continuum map, and are shown only for the area where the extended ionized emission is detected in the radio map at 1420 MHz (see a dotted-dashed box in Figure~\ref{fig1}a). 
The kinematic distance of each molecular clump was also computed by \citet{heyer01}, and the mean kinematic distance 
is found to be $\sim$4.1 kpc. The distance of $\sim$4.1 kpc was adopted for some known star-forming regions 
\citep[i.e., AFGL 5090, AFGL 437, IRAS 03063+5735;][]{Snell1988,Harju1998,Kobulnicky2012}. \citet{Du2017} adopted a distance of $\sim$2.1 kpc for the GFMC Grand Canal located in the Perseus Arm, and a distance of $\sim$3 kpc was used by \citet{Green1989b}. On the other hand, \citet{Karr2003} assumed a distance of $\sim$2.0 kpc to both the LBN sources, which is a similar distance and V$_{\rm lsr}$ (i.e., $-$39 km s$^{-1}$) of nearby W3/W4/W5 regions \citep[e.g.,][]{Karr2003b,navarete19}. Hence, following the work of \citet{Karr2003} and \citet{Du2017}, we consider the distance of $\sim$2 kpc to both the LBN sources in this work. 

Based on the previous published works, we find that the identification of embedded filaments and their involvement in the ongoing physical mechanisms are yet to be performed in the GFMC hosting both the LBN sources, where the existence of high-mass OB-stars has been reported. In this relation, we examine the IR, sub-mm, and molecular maps to study the physical environments of LBN 140.07+01.64 and LBN 140.77$-$1.42. The high resolution near-infrared (NIR) photometric data have been employed to study the distribution of YSOs. Furthermore, we also explore the possible physical process responsible for the existence of the extended ionized nebulae, which has not yet been thoroughly investigated. 

The multi-wavelength data sets are introduced in Section~\ref{sec:data}. 
The study of the embedded filaments and physical environment of the target site is performed in Section~\ref{sec:results}. 
Section~\ref{sec:disc} covers the implications of our observed findings. 
Section~\ref{sec:conc} summarizes the main results of our study.
\section{Data sets}
\label{sec:data}
The present work deals mainly with a target area of $\sim$1$\degr$.96 $\times$ 4$\degr$ (central coordinates: {\it l} = 140$\degr$.55; {\it b} = $-$0$\degr$.461) in the direction of the extended ionized emission hosting LBN 140.07+01.64 and LBN 140.77$-$1.42 (see a dotted-dashed box in Figure~\ref{fig1}a). 
We studied observational data sets collected from different Galactic plane surveys, which are listed in Table~\ref{tab1}. 

We also examined the {\it Herschel} temperature and column density maps (resolution $\sim$12$''$), 
which are not available for the entire selected target area. 
Using the {\it Herschel} continuum images at 70--500 $\mu$m \citep{Molinari10a}, the {\it Herschel} temperature and column density maps were produced for the {\it EU-funded ViaLactea project} \citep{Molinari10b}. 
The Bayesian {\it PPMAP} procedure \citep{marsh15,marsh17} was adopted to produce these {\it Herschel} maps.
\begin{table*}
\scriptsize
\setlength{\tabcolsep}{0.1in}
\centering
\caption{List of different surveys studied in this paper.}
\label{tab1}
\begin{tabular}{lcccr}
\hline 
  Survey  &  Wavelength/Frequency/line(s)       &  Resolution         &  Reference \\   
\hline
\hline 
Canadian Galactic Plane Survey (CGPS)                   &1.42 GHz &   1$'$ $\times$ 1$'$csc$\delta$  &\citet{Taylor2003}\\
NRAO VLA Sky Survey (NVSS)                              & 1.4 GHz        & $\sim$45$''$   &\citet{Condon1998}\\
CGPS $^{12}$CO(J =1--0)   &2.6 mm  & $\sim$100\rlap.{$''$}4   &\citet{Taylor2003}\\ 
{\it Herschel} Infrared Galactic Plane Survey (Hi-GAL) & 160 $\mu$m & 12$''$ &\citet{Molinari10a}\\
Wide-field Infrared Survey Explorer (WISE)               &12 $\mu$m &$\sim$6\rlap.{$''$}5 & \citet{Wright2010}\\
UKIDSS Galactic Plane Survey (GPS) &1.25--2.2 $\mu$m &$\sim$0\rlap.{$''$}8  &\citet{Lawrence2007}\\
Two Micron All Sky Survey (2MASS)     &1.25--2.2 $\mu$m  & $\sim$2\rlap.{$''$}5 &\citet{Skrutskie2006}\\
\hline          
\end{tabular}
\end{table*}
\section{Results}
\label{sec:results}
\subsection{Extended ionized Y-shaped feature and its multi-wavelength view}
\label{sec:yfeature}
\begin{figure*}
\includegraphics[width=\textwidth]{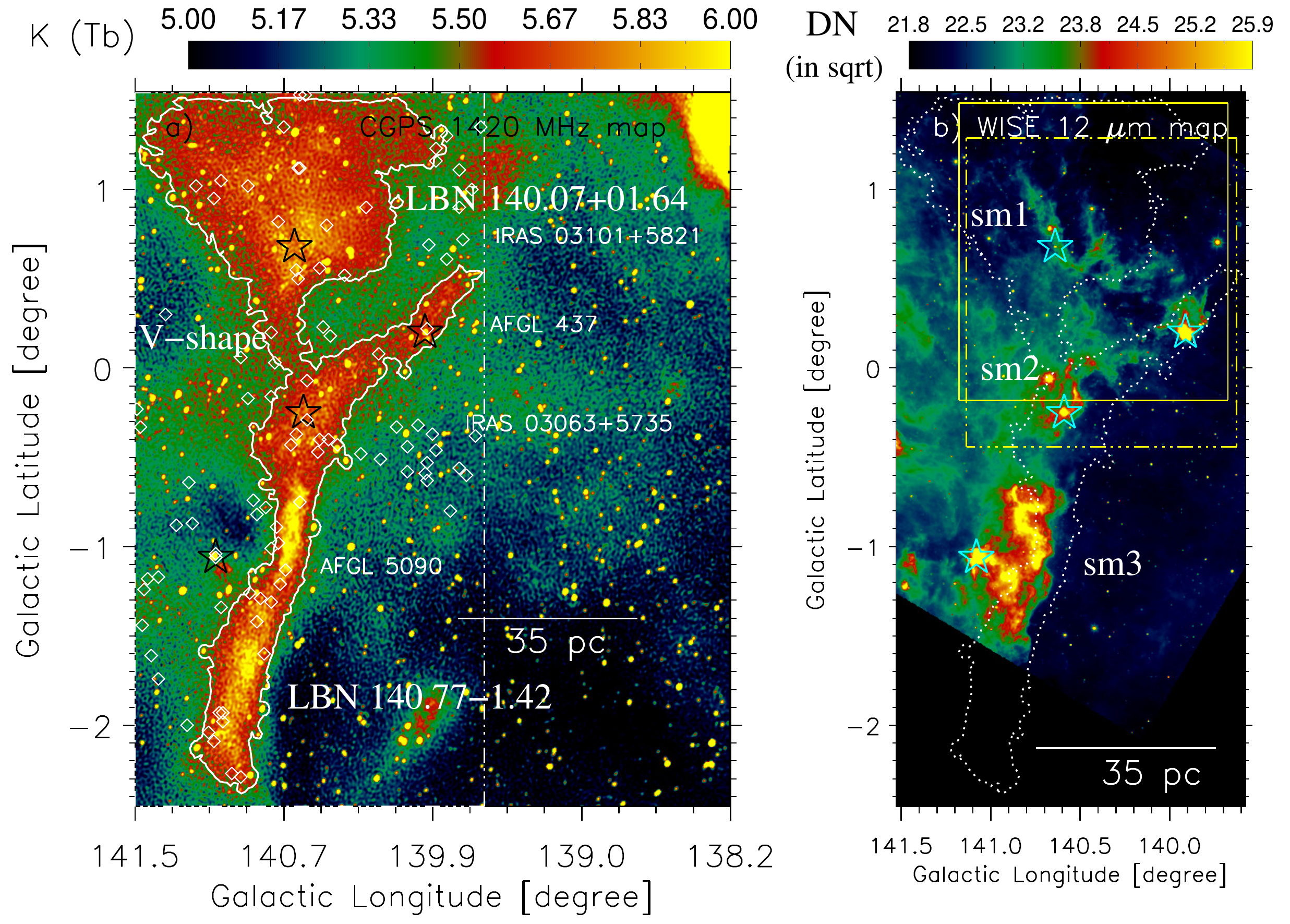}
\caption{a) The panel shows the CGPS 1420 MHz continuum image (area $\sim$3$\degr$.35 $\times$ 4$\degr$; central coordinates: {\it l} = 139$\degr$.86; {\it b} = $-$0$\degr$.469). The CGPS 1420 MHz continuum emission contour at 5.5 K \citep[1$\sigma$ $\sim$71 sin $\delta$ mK;][]{Taylor2003} traces the Y-feature. The positions of molecular clumps \citep[from][]{heyer01} are also shown by diamonds. 
b) WISE 12 $\mu$m image (area $\sim$1$\degr$.96 $\times$ 4$\degr$; central coordinates: {\it l} = 140$\degr$.55; {\it b} = $-$0$\degr$.461) in the direction of Y-feature (see a dotted-dashed box in Figure~\ref{fig1}a). 
A dotted-dashed box highlights an area presented in Figure~\ref{fig2}a, while a solid box encompasses an area shown in 
Figure~\ref{fig2}b. A dotted contour outlines the Y-feature as shown in Figure~\ref{fig1}a. 
In each panel, star symbols indicate the positions of different star-forming regions, 
and the scale bar referring to 35 pc (at a distance of 2.0 kpc) is shown.}
\label{fig1}
\end{figure*}

In the direction of our selected target area, the CGPS 1420 MHz continuum image has allowed us to identify an extended Y-shaped morphology (hereafter, Y-feature; linear extent $\sim$3$\degr$.7 or $\sim$129 pc at a distance of 2 kpc), which consists of a linear part and a V-like structure (see Figure~\ref{fig1}a). 
The Y-feature is traced with the CGPS 1420 MHz continuum emission contour level of 5.5~K \citep[1$\sigma$ $\sim$71 sin $\delta$ mK;][]{Taylor2003}. 
The site LBN 140.77$-$1.42 is spatially seen toward the linear part of the Y-feature, while the sites LBN 140.07+01.64 and AFGL 437 are located toward the opposite sides of the V-like structure. 
We also find the spatial distribution of molecular clumps at [$-$43, $-$34] km s$^{-1}$ \citep[from][]{heyer01} toward the entire Y-feature, confirming its physical existence (see also Section~\ref{sec:mmaps}). The previously reported GFMC Grand Canal \citep[e.g.,][]{Du2017} is seen toward the ionized Y-feature. However, the site AFGL 437 is spatially located away from the GFMC.  

Figure~\ref{fig1}b shows the WISE image at 12 $\mu$m of an area containing the Y-feature, displaying the presence of warm dust emission and large-scale filaments extended toward both the LBN sources. 
In our selected target area, the WISE map is available only for a latitude range of [$-$1$\degr$.6, 1$\degr$.5]. 
Based on the visual inspection of the WISE image, three subregions, sm1, sm2, and sm3 are chosen toward the northern, central, and southern parts of the Y-feature, respectively. In the WISE image, the southern part (i.e., sm3) of the Y-feature is seen with higher intensity than the northern and central parts. 

\begin{figure*}
\includegraphics[width=10.5cm]{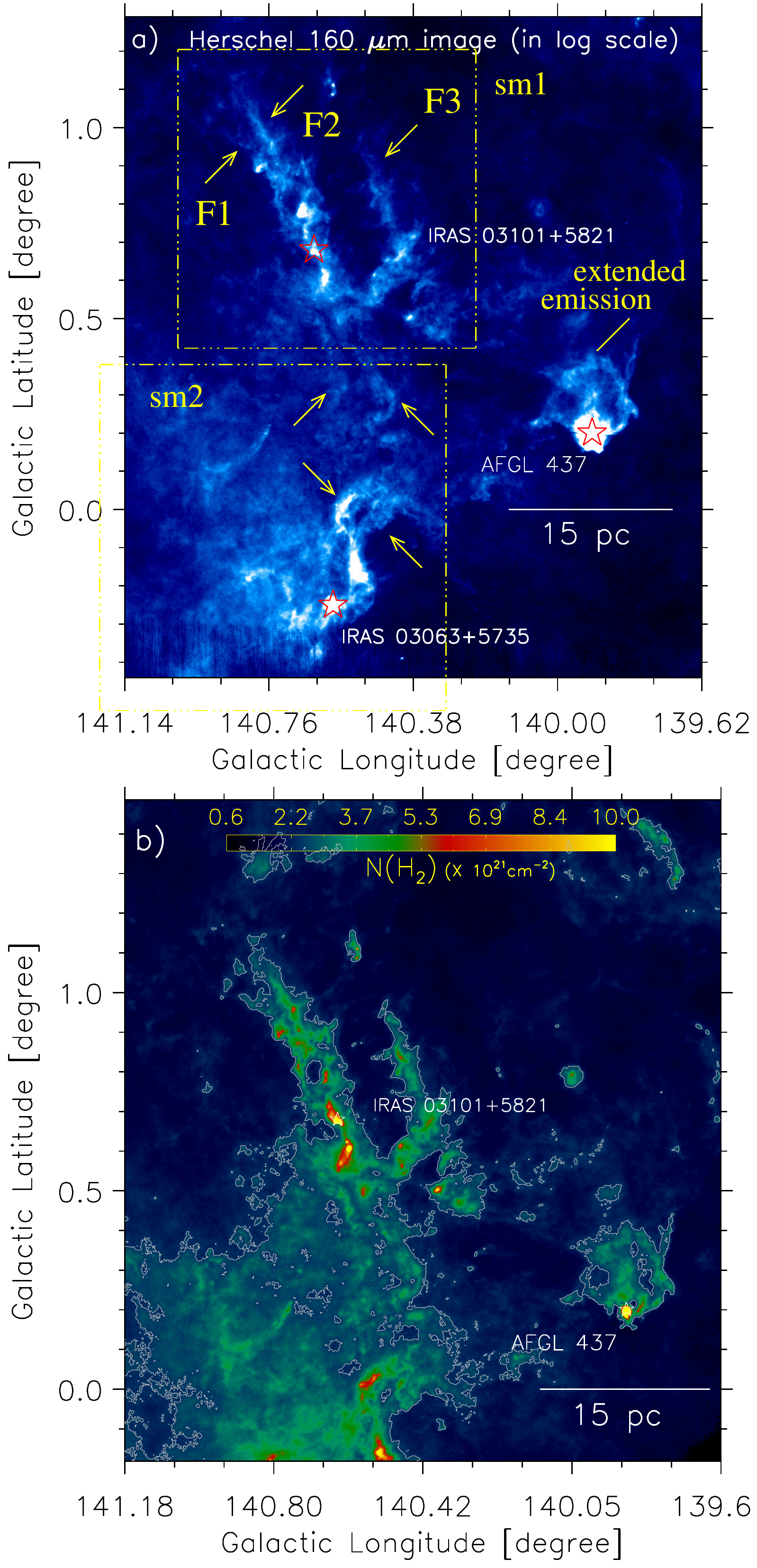}
\caption{a) {\it Herschel} continuum image at 160 $\mu$m of an area highlighted by a dotted-dashed box in Figure~\ref{fig1}b. 
Arrows highlight embedded filaments. Two subregions (i.e., sm1 and sm2) are highlighted by dotted-dashed boxes. 
b) {\it Herschel} column density (N(H$_{\rm 2}$)) map of an area highlighted by a solid box in Figure~\ref{fig1}b. 
The column density contour (in gray) is also shown with a level of 2.4 $\times$10$^{21}$~cm$^{-2}$. 
In each panel, star symbols are the same as in Figure~\ref{fig1}a.}
\label{fig2}
\end{figure*}

To further examine the embedded filaments, Figure~\ref{fig2}a presents the {\it Herschel} image at 160 $\mu$m toward an area hosting the sites AFGL 437, IRAS 03101+5821, and IRAS 03063+5735 (see a dotted-dashed box in Figure~\ref{fig1}b). The areas of two subregions (i.e., sm1 and sm2) are also indicated in Figure~\ref{fig2}a. Note that the {\it Herschel} observations partially cover the area of subregion sm2, 
and do not cover the southern part (i.e., sm3) of the Y-feature. 
At least three filaments (i.e., F1, F2, and F3) are spatially seen toward the subregion sm1, and 
two of them (i.e., F1 and F2) are also extended toward the subregion sm2 (see arrows in Figure~\ref{fig2}a). 
We also see overlapping areas of the filaments F1 and F2 in the {\it Herschel} image at 160 $\mu$m. 
The quantitative information of the embedded features can be examined in Figure~\ref{fig2}b. The site AFGL 437 is associated with the extended 160 $\mu$m continuum emission, and is known to host an H\,{\sc ii} region \citep[see][]{kumar10}. 
In the direction of AFGL 437, the spatial appearance of the 160 $\mu$m continuum emission is very similar as seen in the {\it Spitzer} image at 8.0 $\mu$m \citep[see Figure~1 in][]{kumar10}. 

Figure~\ref{fig2}b displays the {\it Herschel} column density map \citep[resolution $\sim$12$''$;][]{marsh15,marsh17} of an area highlighted by a solid box in Figure~\ref{fig1}b. The column density contour (N(H$_{\rm 2}$)) at 2.4 $\times$10$^{21}$~cm$^{-2}$ is also shown in Figure~\ref{fig2}b, indicating the presence of higher N(H$_{\rm 2}$) of materials toward the filaments and the site AFGL 437. 
\subsection{Molecular structures}
\label{sec:largescalefilaments}
\subsubsection{Moment-0 and Moment-1 maps}
\label{sec:mmaps}
In this section, we examined the CGPS $^{12}$CO(1--0) line data to study the distribution of molecular gas toward the Y-feature. 
The $^{12}$CO(1--0) emission toward our target area is studied in a velocity range of [$-$42.7, $-$34.4] km s$^{-1}$, allowing us to generate the integrated intensity (or moment-0) map and the intensity weighted mean velocity (or moment-1) map. For the moment-1 map, the molecular emission is clipped with a value of 3.9 K.

\begin{figure*}
\includegraphics[width=\textwidth]{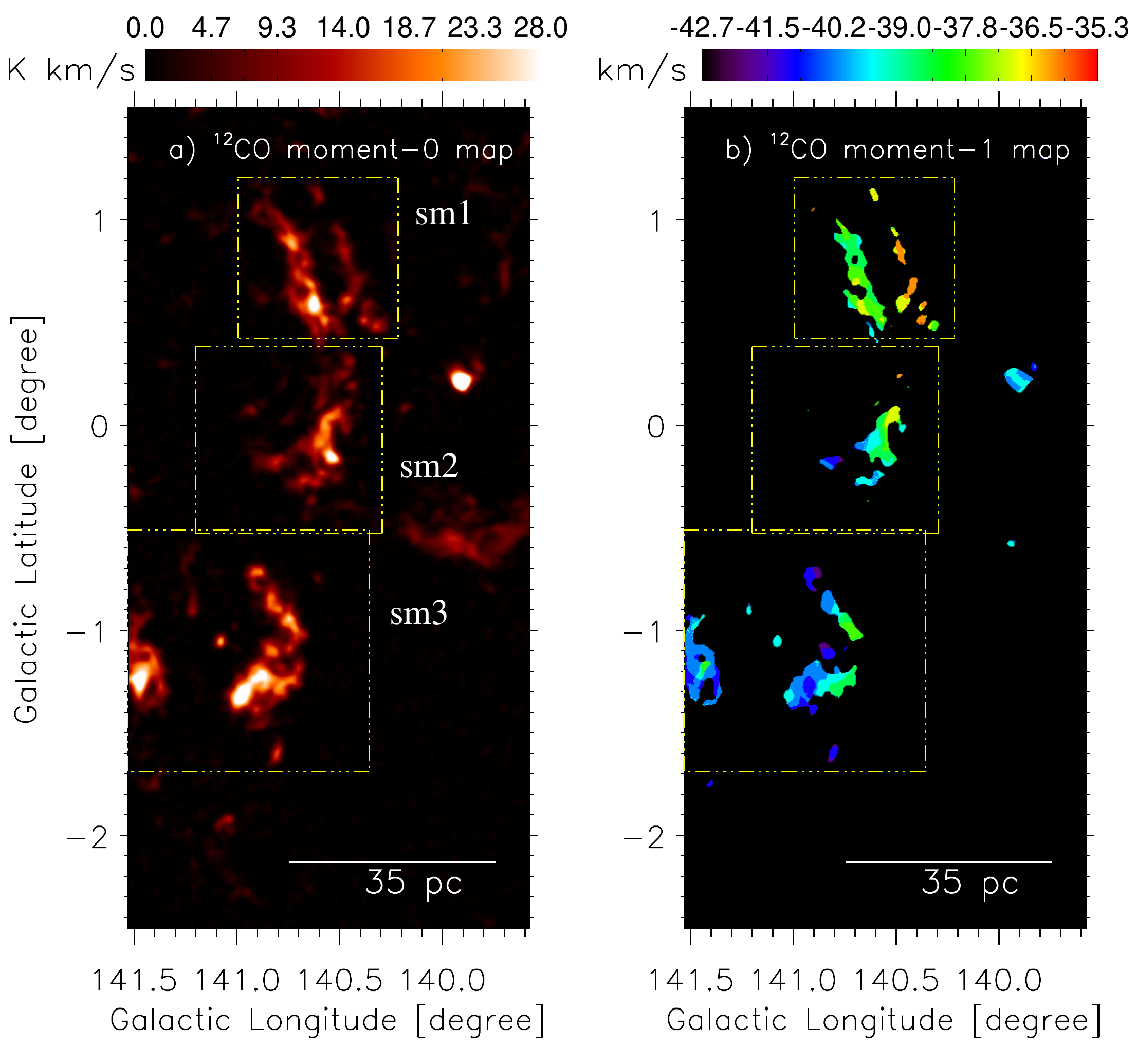}
\caption{a) The CGPS $^{12}$CO(1--0) molecular emission map (moment-0 map) toward the Y-feature. 
The molecular emission is integrated over a velocity range of [$-$42.7, $-$34.4] km s$^{-1}$. 
b) The $^{12}$CO(1--0) moment-1 map. 
In each map, three subregions sm1, sm2, and sm3 are indicated by dotted-dashed boxes.} 
\label{fig3}
\end{figure*}

The moment-0 and moment-1 maps of the $^{12}$CO(1--0) emission are presented in Figures~\ref{fig3}a and~\ref{fig3}b, respectively. 
The areas of three subregions (i.e., sm1, sm2, and sm3) are also marked in the moment-0 map, and are located toward the previously reported GFMC Grand Canal \citep[i.e., a continuous elongated structure; see Figure~10 in][]{Du2017}. 
In the direction of sm3, a compact gas condensation is traced around AFGL~5090, which is surrounded by two arc-like molecular features. The arc-like molecular feature located toward the right side of AFGL~5090 
has a filamentary appearance, where the diffuse dust emission is evident in the WISE image. 
In the moment-1 map, the gas velocity difference is evident toward the molecular structures traced in the moment-0 map. 
The blue-shifted molecular gas around $-$41 km s$^{-1}$ is seen toward the subregion sm3, while the red-shifted gas around $-$36 km s$^{-1}$ is depicted toward the subregion sm1 (see Figure~\ref{fig3}b). 
This argument is also supported by the examination of the published position-velocity diagram of molecular gas 
\citep[see Figure~10g in][]{Du2017}. In the direction of AFGL 437, a noticeable velocity variation is also evident, which may suggest the presence of an outflow activity. It is in agreement with the previously reported IR outflow in AFGL 437 \citep{kumar10}. 

\begin{figure*}
\includegraphics[width=16cm]{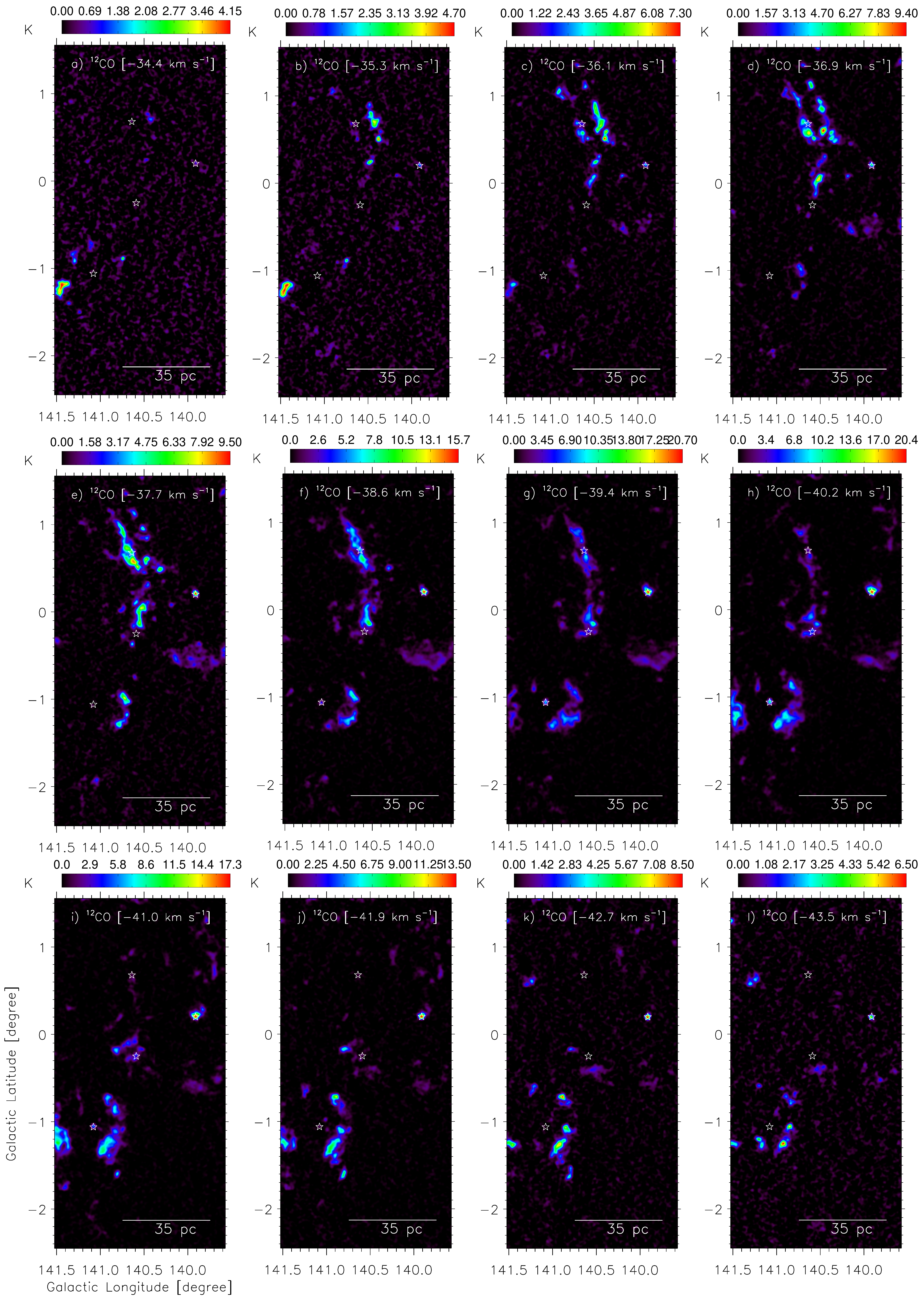}
\caption{Velocity channel maps of the $^{12}$CO(1--0) emission toward the Y-feature. 
The velocity of molecular emission is indicated in each panel (in km s$^{-1}$). 
In each panel, the scale bar and star symbols are the same as in Figure~\ref{fig1}b.} 
\label{fig4}
\end{figure*}

To further explore the distribution of molecular gas, in Figure~\ref{fig4}, we present velocity channel maps of $^{12}$CO(1--0), where the positions of star-forming sites are also marked by star symbols. 
Figure~\ref{fig4} confirms the existence of a continuous elongated molecular structure in the direction of the Y-feature. 
The molecular gas is evident toward the {\it Herschel} dust filaments (i.e., F1, F2, and F3; see Figure~\ref{fig2}a). 
It also appears that the filaments F1 and F2 are large scale molecular structures, and are present in all the 
subregions (i.e., sm1, sm2, and sm3; see Figures~\ref{fig4}c--\ref{fig4}f). 
It is interesting to note that Figure~\ref{fig4}f traces the two curved and coupled filaments (i.e., F1 and F2) in the northern direction.

\begin{figure*}
\includegraphics[width=\textwidth]{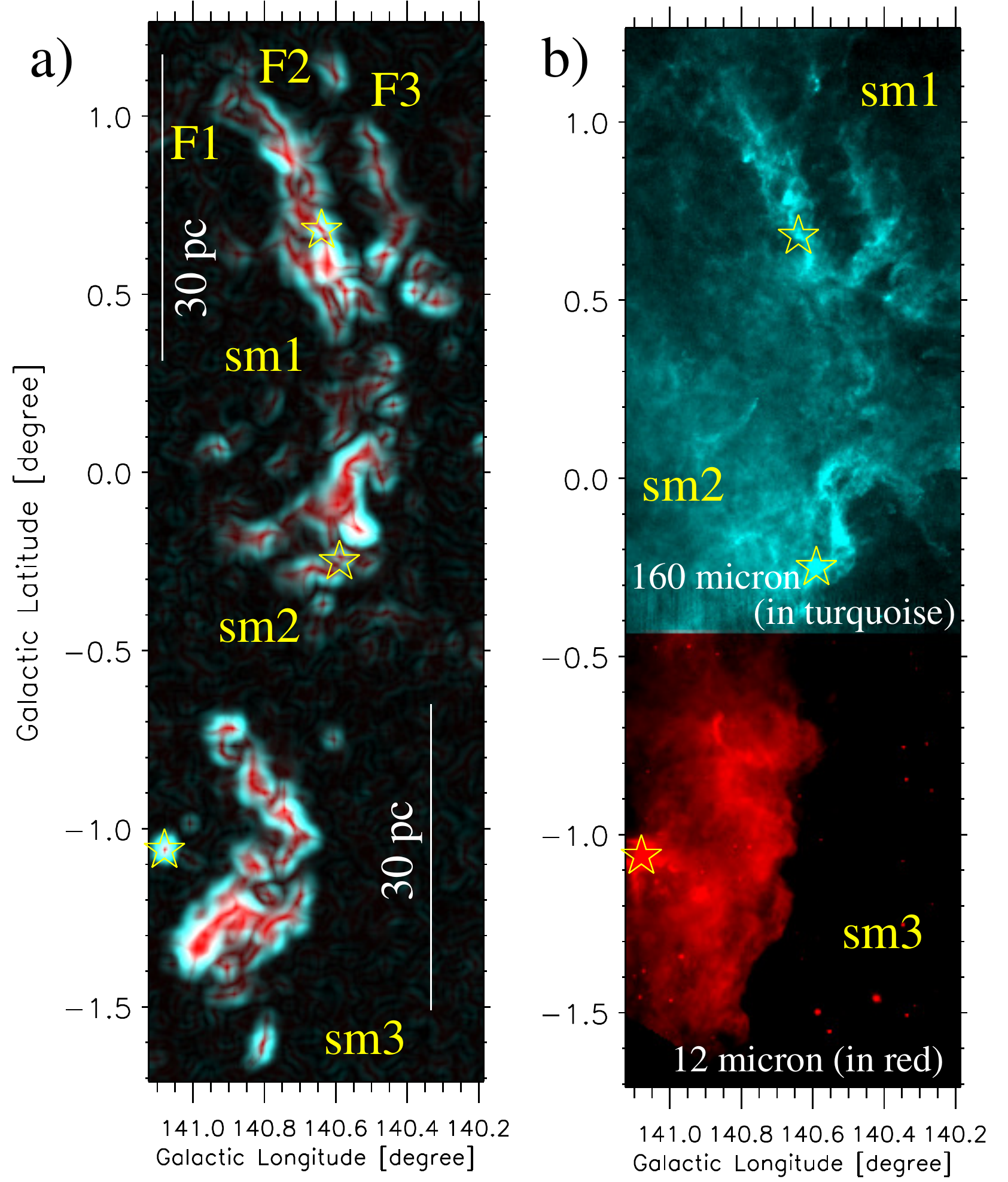}
\caption{a) The panel shows a two-color composite image made using the $^{12}$CO(1--0) moment-0 map. 
In the color composite image, the ``sobel" processed $^{12}$CO(1--0) moment-0 map is shown in red color, 
while the $^{12}$CO(1--0) moment-0 map is displayed in turquoise color. 
The sobel filter is employed for edge detection \citep[see text for more details and also][]{sobel14}.  
b) The panel presents the infrared image toward Y-feature. 
The {\it Herschel} 160 $\mu$m image (in turquoise color) is shown toward subregions sm1 and sm2, while the WISE 12 $\mu$m image (in red) is displayed toward the subregion sm3. 
In each panel, star symbols are the same as in Figure~\ref{fig1}a.}
\label{fig5}
\end{figure*}

To spatially outline the molecular filaments, Figure~\ref{fig5}a presents a two-color composite map (red: {\it IDL} based routine ``sobel" processed moment-0 map, turquoise: moment-0 map) of an area hosting all the subregions (i.e., the GFMC Grand Canal) excluding the site AFGL 437. The sobel filter is often utilized for edge detection \citep{sobel14}, allowing us to uncover the molecular filaments in the Y-feature. 
The sobel filter finds the edges by seeking for the minimum and maximum in the first derivative of the input image \citep{sobel14}. 
For a comparison, Figure~\ref{fig5}b shows the IR image of the same area as seen in Figure~\ref{fig5}a. 
In Figure~\ref{fig5}b, the {\it Herschel} 160 $\mu$m image (in turquoise color) covers the subregions sm1 and sm2, and the WISE 12 $\mu$m image (in red) is shown toward the subregion sm3.
In Figure~\ref{fig5}a, the filaments F1 and F2 seem to be coupled to each other, revealing a twisted behaviour. 
However, the filament F3 appears as a separate branch. 
To further confirm the possible twisted nature of the filaments, the knowledge of the velocity structure along each filament is required.  
The coupled filamentary features can also be seen in the direction of subregions sm2 and sm3, but they are not as prominent as seen in the subregion sm1 (see also Figure~\ref{fig5}b). 
\subsubsection{Velocity structures of the filaments}
\label{subsec:velfil}
\begin{figure*}
\includegraphics[width=\textwidth]{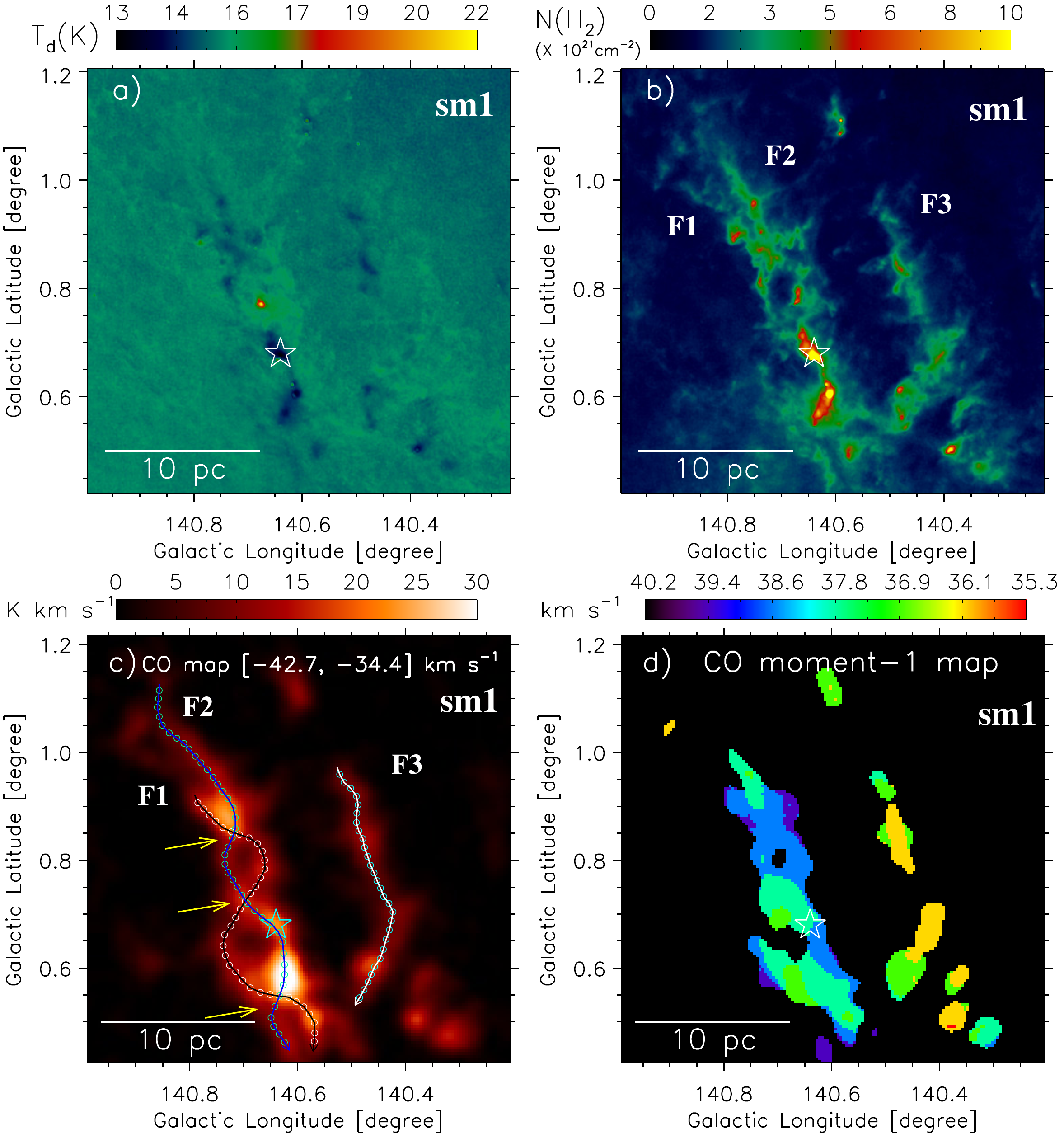}
\caption{a) {\it Herschel} temperature map of the subregion ``sm1" (see a dotted-dashed box in Figure~\ref{fig3}a).
b) {\it Herschel} column density map of ``sm1". c) The $^{12}$CO(1--0) emission (moment-0) map of ``sm1" (see Figure~\ref{fig3}a). 
Three Filaments (i.e., F1, F2, and F3) are highlighted by curves. 
Several small open circles (radii = 20$''$) are also marked along each filament, and an average spectrum over each circle is examined. Arrows highlight the common zones of the filaments F1 and F2. d) The $^{12}$CO(1--0) moment-1 map of ``sm1" (see Figure~\ref{fig3}b). 
In each panel, the star symbol is the same as in Figure~\ref{fig1}a.} 
 \label{fig7}
\end{figure*}

To further explore the filaments, F1--F3, a zoomed-in view of the area containing the subregion sm1 is presented in Figure~\ref{fig7}.  
Figures~\ref{fig7}a and \ref{fig7}b present the {\it Herschel} temperature and column density maps of the subregion sm1, respectively. 
In Figures~\ref{fig7}c and~\ref{fig7}d, we show the $^{12}$CO moment-0 map at [$-$42.7, $-$34.4] km s$^{-1}$ and the moment-1 map of the subregion sm1, respectively. In the {\it Herschel} maps, we find the filaments (i.e., F1, F2, and F3) with T$_{\rm d}$ of $\sim$13--15~K and N(H$_{\rm 2}$) $\sim$2.4--10 $\times$10$^{21}$~cm$^{-2}$. Each molecular filament is highlighted by a curve in Figure~\ref{fig7}c, where several small circles (radii $\sim$20$''$) are also marked. 
We examined an average spectrum over each circle, and obtained a peak velocity through the fitting of the observed spectrum with a Gaussian function. 
The star-forming site IRAS 03101+5821 is located toward either filament F1 or F2. 
In Figure~\ref{fig7}d, one can also find the noticeable gas velocity difference of about 2-3 km s$^{-1}$ toward the filaments F1 and F2. 
The filament F3, which appears as an adjacent filamentary branch, also shows considerable velocity difference.

\begin{figure*}
\includegraphics[width=13cm]{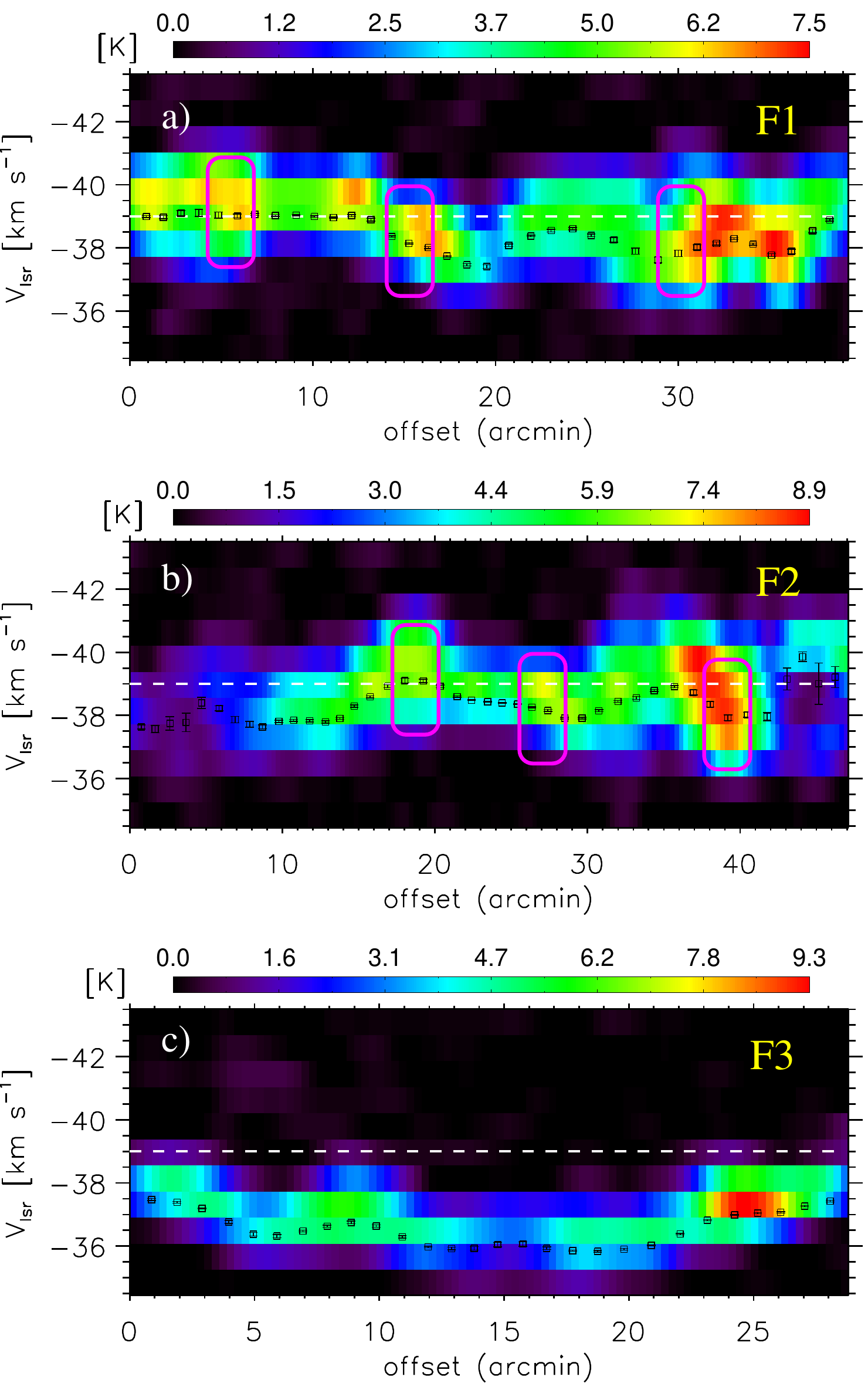}
\caption{Position-velocity diagrams along the filaments a) ``F1", b) ``F2", and c) ``F3" as indicated in Figure~\ref{fig7}c. 
In panels ``a" and ``b", pink boxes represent the common zones of the filaments F1 and F2 (see also arrows in Figure~\ref{fig7}c). In all panels, black squares show mean velocities (along with error bars) derived through the Gaussian fitting of averaged spectra over circular regions (see small open circles in Figure~\ref{fig7}c). 
In each panel, one arcmin (1$'$) corresponds to 0.58 pc (at a distance of 2 kpc), 
and a horizontal dashed line (in white) is marked at V$_\mathrm{lsr}$ = $-$39 km s$^{-1}$.}
 \label{fig8}
\end{figure*}

Figures~\ref{fig8}a,~\ref{fig8}b, and~\ref{fig8}c present the position-velocity diagrams along the filaments F1, F2, and F3, respectively (see curves shown in Figure~\ref{fig7}c). The majority of the molecular gas in the direction of the filaments F1 and F2 is traced in a velocity range of [$-$36, $-$41] km s$^{-1}$, while the majority of the molecular gas toward the filament ``F3" is depicted in a velocity range of [$-$35, $-$39] km s$^{-1}$. In Figures~\ref{fig8}a and~\ref{fig8}b, the common zones of the filaments F1 and F2 are 
highlighted by pink boxes (see also arrows in Figure~\ref{fig7}c). 
We find a velocity gradient of about 0.3--0.5 km s$^{-1}$ pc$^{-1}$ toward these common zones. 
The common zones of the filaments are also associated with a relatively high intensity compared to their other parts. 
From Figures~\ref{fig7}c and~\ref{fig8}, the spatial extension of the common zones is found to be about 6$'$--8$'$ (or $\sim$3.5--4.6 pc). The observed high intensities and velocity gradient toward the common zones suggest the possible interaction between both the filaments. 
However, we do not find such an intensity variation along the filament F3. 

The coarse beam of the CGPS $^{12}$CO(1--0) line data does not allow us to get more insights into the velocity structures of the filaments. 
However, based on a visual examination, there is a hint of an oscillatory-like velocity pattern along both the filaments F1 and F2, 
showing velocity variations. 
To further examine this aspect, we display mean velocities (along with error bars) of averaged spectra over circular 
regions as highlighted along each filament (see circles in Figure~\ref{fig7}c). 
This particular analysis indicates the presence of the oscillatory-like velocity pattern along the filaments F1 and F2, but it is not clearly seen in the case of F3. 
Furthermore, this velocity pattern is more obvious in the case of F2 compared to F1. The implication of this finding is discussed in Section~\ref{sec:disc}. 
\subsection{Star formation activities}
\label{sec:largescaleyso}
\begin{figure*}
\includegraphics[width=\textwidth]{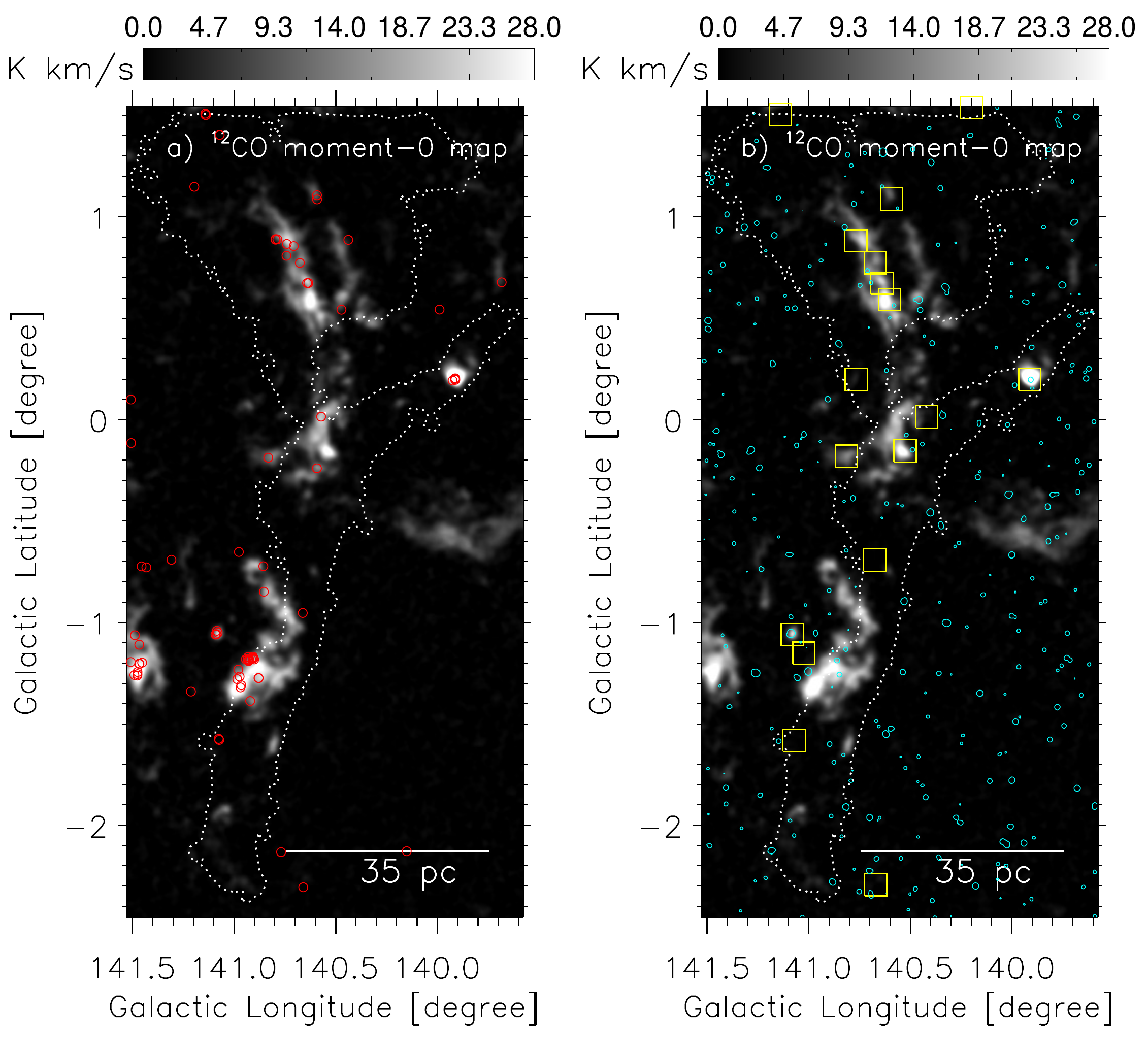}
\caption{a) Overlay of the positions of infrared-excess sources (with H$-$K $>$ 1.3 mag; see open circles) 
on the $^{12}$CO(1--0) emission map. 
b)  Overlay of the locations of YSOs presented in \citet{Karr2003} (see open squares) and 
the NVSS 1.4 GHz continuum emission contour on the $^{12}$CO(1--0) emission map.
The NVSS contour (in cyan) is also shown with a level of 5$\sigma$, where 1$\sigma$ = 0.45 mJy beam$^{-1}$.
In both panels, the molecular map is the same as shown in Figure~\ref{fig3}a.
In each panel, the dotted contour showing the Y-feature and the scale bar are the same as in Figure~\ref{fig1}b.}
\label{fig6}
\end{figure*}

To infer the infrared-excess sources/YSOs toward the Y-feature, we examined a color-magnitude (H$-$K/K) diagram of point-like sources detected in the H and K bands. In this relation, we used the reliable H- and K-band photometric data from the UKIDSS-GPS and 2MASS surveys. Detailed procedures for the selection of H and K sources are given in \citet{lucas08} and \citet{dewangan15,dewangan17}. A color condition of H$-$K $>$ 1.3 mag is chosen to identify the YSO candidates, and is found through the study of the color-magnitude space of a nearby control field.
A total of 80 YSO candidates are selected in the direction of Y-feature. 
In Figure~\ref{fig6}a, the positions of the selected YSO candidates are overlaid on the moment-0 map of $^{12}$CO (see circles). A large number of YSOs are seen toward the site AFGL 5090, while the site AFGL 437 hosts a few YSOs. 
Figure~\ref{fig6}b displays the locations of previously known YSOs (see squares) overlaid on the $^{12}$CO map, which were identified using the IRAS color conditions \citep[from][]{Karr2003}. 
The NVSS 1.4 GHz continuum emission contours are also overlaid on the $^{12}$CO map, allowing us to infer the locations of the H\,{\sc ii} regions toward the Y-feature.

We find ongoing star formation activities toward the entire Y-feature, which hosts LBN 140.07+01.64, LBN 140.77$-$1.42, and some previously reported star-forming sites/H\,{\sc ii} regions (i.e., IRAS 03035+5819/AFGL~437, AFGL~5090, IRAS 03101+5821, and IRAS 03063+5735). From Figures~\ref{fig6}a and~\ref{fig6}b, noticeable YSO candidates are also inferred toward the common zones of the filaments F1 and F2, where high column densities are depicted in the {\it Herschel} column density map (see Figure~\ref{fig2}b).  

Taking all these results together, high-mass OB-stars and YSOs are present in the Y-feature. 
\section{Discussion}
\label{sec:disc}
We identified a large-scale ionized Y-feature (linear extent $\sim$3$\degr$.7 or $\sim$129 pc at a distance of 2 kpc) located in the Perseus Arm (see Figure~\ref{fig1}a).
Furthermore, the IR, sub-mm, and molecular line data have revealed at least two large-scale filaments (F1 and F2) in the Y-feature (see Section~\ref{sec:largescalefilaments}). The linear extension of each filament is more than 20 pc. The spatial appearance of these filaments hints their coupling nature, revealing several possible overlapping zones of high column densities. 
The filamentary twisting/coupling appears more significant in the northern and central areas of the Y-feature. 
The initial assessment of twisting/coupling of the filaments is performed on the two-dimensional (2D) images, but a twisted shape is a three-dimensional (3D) morphology. Therefore, we have studied the velocity structures of the filaments using 
the CGPS $^{12}$CO line data (see Section~\ref{subsec:velfil}), which have a coarse beam size (see Table~\ref{tab1}). In this relation, the position-velocity diagrams along the filaments are examined, and reveal a signature of an 
oscillatory-like velocity pattern along each filament, separately (see Figures~\ref{fig8}a and~\ref{fig8}b). 
Note that in this work, we are dealing with two spatially coupled filaments. Hence, in general, one may expect such an oscillation-like velocity pattern, which may be related to the possible twisted behaviour. 
However, a physical model is required to successfully explain the observed spatial 
and velocity structures, and such a study is beyond the scope of this work. 

Concerning this underlying velocity structure, in the literature, we find a promising filament G350.5-N associated with the cloud G350.54+0.69, where a large-scale periodic velocity oscillation was 
reported by \citet{liu19} using the $^{13}$CO(2--1) and C$^{18}$O(2--1) line data (resolution $\sim$28$''$; see Figures~6 and 7 in their paper). This observed velocity structure was suggested to be originated from the combined effects of core/fragment formation \citep[e.g.,][]{kainulainen16,dewangan19} and the large-scale physical oscillation along the filament \citep[see][for more details]{liu19}. Based on our study presented in this paper, we suggest that the observed velocity structure along the filaments, F1 and F2, may be related to large-scale physical oscillations, and hints their possible twisted behaviour. 
High-resolution continuum and molecular line observations will be helpful to further explore the twisted filaments in the Y-feature.
Considering these results, we suggest that it is the first observational work reporting the possible twisting of the large-scale molecular and dust filaments, which is the most striking outcome of this work.

The involvement of embedded filaments in star formation processes (including low- and high-mass stars) has been known in the literature. 
In this relation, signposts of star formation activities are largely investigated at the intersection/merging/collision zones of filaments as well as the junction of multiple parsec scale filaments \citep[i.e., hub-filament system;][]{myers09,motte18}. 
Most recently, a theoretical study on merging filaments has been carried out \citep[][submitted to MNRAS]{Hoemann21}. \citet{Hoemann21} studied the merging behavior of parallel aligned filaments. On the other hand, in the case of isolated filaments, the end-dominated collapse (EDC) process has been proposed, and favours the fragmentation and collapse at the ends of the filaments \citep{Pon12,clarke15}. 
However, the twisting of filaments is an uncommon phenomenon, and has not been introduced in the literature so far. 
In systems hosting the twisting of filaments, one may expect multiple overlapping areas of high column density regions, where star formation may proceed. Hence, the follow up theoretical studies are highly essential to explore the role of twisting and coupling of filaments in star formation processes.  

We find the presence of high-mass OB-stars and YSOs in the direction of the Y-feature (see Section~\ref{sec:largescaleyso}), showing signs of ongoing star formation. In particular, in the northern part of the Y-feature, noticeable YSOs are found toward the possible common zones of the filaments, F1 and F2 (see circles in Figure~\ref{fig6}a and squares in Figure~\ref{fig6}b). 
On a global scale of the Y-feature, the star formation seems to be guided by the large-scale coupling and twisted nature of filaments. 

Previously, it was proposed that LBN 140.77$-$1.42 may be a large-scale ionization front penetrating a large interstellar cloud \citep{Green1989b}. Later, \citet{Karr2003} suggested that the ionized material in both the LBN sources could be illuminated by the emission from a nearby high-mass O-star HD~16691 and the ultra-violet (UV) photons from W5 and W3/W4 regions. Hence, it is possible that the origin of the large-scale Y-feature may be influenced by the energetics of high-mass stars (i.e., ionized emission, stellar wind, and radiation pressure), but stellar feedback from the nearby high-mass stars/complexes may not be the only mechanism to explain the origin of the large-scale Y-feature.
Some other physical process may be ongoing in the Y-feature, which could explain its origin. In this context, the role of the molecular and dust filaments is very promising.  
We suggest that the large-scale coupling and twisting of filaments might have formed the low-mass and high-mass stars.
Additionally, the influence of the previously known high-mass stars to their immediate surroundings cannot be ruled out. 

Overall, our proposed physical process has a significant contribution to explain the existence of the ionized Y-feature and the observed star formation activities.
\section{Summary and Conclusions}
\label{sec:conc}
This paper uses multi-wavelength and multi-scale data to study morphology and kinematics of the gas toward the two Lynds Bright Nebulae, LBN 140.07+01.64 and LBN 140.77$-$1.42. Both the LBN sources are found to be a part of a large scale ionized Y-feature (linear extent $\sim$3$\degr$.7 or $\sim$129 pc at a distance of 2 kpc) that is identified in the CGPS 1420 MHz continuum map. The previously identified star-forming sites/H\,{\sc ii} regions (i.e., IRAS 03035+5819/AFGL~437, AFGL~5090, IRAS 03101+5821, and IRAS 03063+5735) are spatially seen in the direction of Y-feature, hosting high-mass OB-stars. 
Infrared-excess sources/YSOs are found in the direction of the entire Y-feature, enabling us to infer the ongoing star formation activities.
The inspection of the IR and sub-mm images reveals at least two large-scale dust filaments extended toward both the LBN sources.   
In the direction of the northern and central parts of the Y-feature, the {\it Herschel} column density and temperature maps display higher column density ($\ge$ 2.4 $\times$10$^{21}$~cm$^{-2}$) of materials toward the filaments with the dust temperature of $\sim$13--15~K. 
The CGPS $^{12}$CO(1--0) line data depict the large-scale filamentary structures toward the Y-feature, which are investigated in a velocity range of [$-$42.7, $-$34.4] km s$^{-1}$. The molecular and dust filaments seem to be twisted and coupled to each other, showing several possible overlapping zones of high column densities. In the northern part of the Y-feature, the study of the velocity structures of the filaments also indicates their possible twisted nature.  

For the first time, the present paper observationally reports the possible twisting/coupling of the large-scale filaments. This underlying mechanism and the energetics of high-mass OB-stars together appear to be responsible for the origin of the ionized Y-feature.
\section*{Acknowledgments}
We thank the anonymous reviewer for several useful comments and 
suggestions, which greatly improved the scientific contents of the paper. 
The research work at Physical Research Laboratory is funded by the Department of Space, Government of India. 
The research presented in this paper has used data from the Canadian Galactic Plane Survey, a Canadian project with international partners, supported by the Natural Sciences and Engineering Research Council. 
This work is based on data obtained as part of the UKIRT Infrared Deep Sky Survey. This publication makes use of data products from the Two Micron All Sky Survey, which is a joint project of the University of Massachusetts and the Infrared Processing and Analysis Center/California Institute of Technology, funded by the National Aeronautics and Space Administration and the National Science Foundation. 
TB acknowledges the support from S. N. Bose National Centre for Basic Sciences under the Department of Science and Technology (DST), Govt. of India. DKO acknowledges the support of the Department of Atomic Energy, Government of India, under project Identification No. RTI 4002.
\subsection*{Data availability}
The CGPS 1420 MHz continuum map and $^{12}$CO(1--0) line data underlying this article are available from the publicly accessible website\footnote[1]{https://www.cadc-ccda.hia-iha.nrc-cnrc.gc.ca/en/jcmt/}.
The NVSS 1.4 GHz continuum data underlying this article are available from the publicly accessible website\footnote[2]{https://www.cv.nrao.edu/nvss/postage.shtml}.
The {\it Herschel}, WISE, and 2MASS data underlying this article are available from the publicly accessible NASA/IPAC infrared science archive\footnote[3]{https://irsa.ipac.caltech.edu/frontpage/}.
The {\it Herschel} column density and temperature maps underlying this article are available from the publicly accessible website\footnote[4]{http://www.astro.cardiff.ac.uk/research/ViaLactea/}.
The UKIDSS GPS data underlying this article are available from the publicly accessible website\footnote[5]{http://wsa.roe.ac.uk/}.
%


\begin{thebibliography}{99}
%
\bibitem[Andr{\'e} et al.(2010)]{andre10}
Andr{\'e} P. et al., 2010, A\&A, 518, L102


%

\bibitem[Clarke \& Whitworth(2015)]{clarke15}
Clarke S.~D., Whitworth A.~P., 2015, MNRAS, 449, 1819
    
\bibitem[Condon et al.(1998)]{Condon1998}
Condon J.~J., Cotton W.~D., Greisen E.~W., Yin Q.~F., Perley R.~A., Taylor G.~B., Broderick J.~J., 1998, AJ, 115, 1693


\bibitem[Dewangan et al.(2015)]{dewangan15} 
Dewangan L.~K., Luna A., Ojha D.~K., Anandarao B.~G., Mallick K.~K., Mayya Y.~D., 2015, ApJ, 811, 79


\bibitem[Dewangan et al.(2017a)]{dewangan17} 
Dewangan L.~K., Ojha D.~K., Zinchenko I., Janardhan P., Luna A., 2017a, ApJ, 834, 22

\bibitem[Dewangan et al.(2017b)]{dewangan17b}
Dewangan L.~K., Baug T., Ojha D.~K., Janardhan P., Devaraj R., Luna A., 2017b, ApJ, 845, 34

\bibitem[Dewangan et al.(2018)]{dewangan18}
Dewangan L.~K., Baug T., Ojha D.~K., Ghosh S.~K., 2018, ApJ, 869, 30

\bibitem[Dewangan et al.(2019)]{dewangan19}
Dewangan L.~K., Pirogov L.~E., Ryabukhina O.~L., Ojha, D. K., Zinchenko I., 2019, ApJ, 877, 1


\bibitem[Dewangan et al.(2020)]{dewangan20x}	
Dewangan L. K., Ojha D. K., Sharma Saurabh, del Palacio S., Bhadari N.~K., Das A., 2020, ApJ, 903, 13

\bibitem[Dewangan(2021)]{dewangan21}	
Dewangan L. K., 2021, MNRAS, 504, 1152 





\bibitem[Digel et al.(1996)]{digel96}
Digel S.~W., Lyder D.~A., Philbrick A.~J., Puche D., Thaddeus P., 1996, ApJ, 458, 561


\bibitem[Du et al.(2017)]{Du2017}
Du X., Xu Y., Yang J., Sun Y., 2017, ApJS, 229, 24


\bibitem[Green(1989)]{Green1989b}
Green D.~A., 1989, AJ, 98, 2210

\bibitem[Harju et al.(1998)]{Harju1998}
Harju J., Lehtinen K., Booth R.~S., Zinchenko I, 1998, A\&AS, 132, 211

\bibitem[Heyer et al.(2001)]{heyer01}
Heyer M.~H., Carpenter J.~H., Snell R.~L., 2001, ApJ, 551, 852

\bibitem[Hoemann et al.(2021)]{Hoemann21}
Hoemann E., Heigl S., Burkert A., 2021, preprint (arXiv210404541)

\bibitem[Kainulainen et al.(2016)]{kainulainen16}
Kainulainen J., Hacar A., Alves J., Beuther H., Bouy H., Tafalla M., 2016, A\&A, 586, 27

\bibitem[Karr \& Martin(2003a)]{Karr2003}
Karr J.~L., Martin, P.~G., 2003a, ApJ, 595, 880

\bibitem[Karr \& Martin(2003b)]{Karr2003b}
Karr J.~L., Martin, P.~G., 2003b, ApJ, 595, 900

\bibitem[Kobulnicky et al.(2012)]{Kobulnicky2012}
Kobulnicky H.~A., Lundquist M.~J., Bhattacharjee A., Kerton C.~R., 2012, AJ, 143, 71

\bibitem[Kumar Dewangan \& Anandarao(2010)]{kumar10} 
Kumar Dewangan L., Anandarao B.~G., 2010 MNRAS 402 2583

\bibitem[Kumar et al.(2020)]{kumar20} 
Kumar M.~S.~N., Palmeirim P., Arzoumanian D., Inutsuka S.~I., 2020, A\&A, 642, A87

\bibitem[Lawrence et al.(2007)]{Lawrence2007}
Lawrence A. et al., 2007, MNRAS, 379, 1599

\bibitem[Liu et al.(2019)]{liu19}
Liu H.-L., Stutz A., Yuan J.-H., 2019, MNRAS, 487, 1259

\bibitem[Lucas et al.(2008)]{lucas08}
Lucas P.~W. et al., 2008, MNRAS, 391, 136

\bibitem[Lynds(1965)]{Lynds1965}
Lynds B.~T., 1965, ApJS, 12, 163

\bibitem[Marsh et al.(2015)]{marsh15} 
Marsh K.~A., Whitworth A.~P., Lomax O., 2015, MNRAS, 454, 4282

\bibitem[Marsh et al.(2017)]{marsh17} 
Marsh K.~A. et al., 2017, MNRAS, 471, 2730

\bibitem[Molinari et al.(2010a)]{Molinari10a}
Molinari S. et al., 2010a, A\&A, 518, L100

\bibitem[Molinari et al.(2010b)]{Molinari10b}
Molinari S. et al., 2010b, PASP, 122, 314

\bibitem[Motte et al.(2018)]{motte18} 
Motte F., Bontemps S., Louvet F., 2018, ARA\&A, 56, 41 

\bibitem[Myers(2009)]{myers09} 
Myers P.~C., 2009, ApJ, 700, 1609

\bibitem[Navarete et al.(2019)]{navarete19}
Navarete, F., Galli, P.~A.~B., Damineli, A., 2019, MNRAS, 487, 2771

\bibitem[Pon et al.(2012)]{Pon12}
Pon A., Toal{\'a} J.~A., Johnstone D., V{\'a}zquez-Semadeni E., Heitsch F., G{\'o}mez G.~C., 2012, ApJ, 756, 145

\bibitem[Skrutskie et al.(2006)]{Skrutskie2006}
Skrutskie M.~F. et al., 2006, AJ, 131, 1163

\bibitem[Snell et al.(1988)]{Snell1988}
Snell R.~L., Huang, Y.~-L., Dickman R.~L., Claussen M.~J., 1988, ApJ, 325, 853

\bibitem[Sobel(2014)]{sobel14}
Sobel, I., 2014. An Isotropic 3x3 Image Gradient Operator. Presentation at Stanford A.I, Project, p. 1968 
(link: http://refhub.elsevier.com/S0892-6875(21)00057-1/h0260)

\bibitem[Taylor et al.(2003)]{Taylor2003}
Taylor A.~R. et al., 2003, AJ, 125, 3145

\bibitem[Tig{\'e} et al.(2017)]{Tige+2017} 
Tig{\'e} J. et al., 2017, A\&A, 602, A77 

\bibitem[Trevi{\~n}o-Morales et al.(2019)]{morales19}
Trevi{\~n}o-Morales S.~P. et al., 2019, A\&A, 629, A81

\bibitem[Wright et al.(2010)]{Wright2010}
Wright E.~L. et al., 2010, AJ, 140, 1868


%
\end{thebibliography}
\end{document}